\documentclass[12pt,oneside,reqno]{amsart}	
\usepackage[margin=1in]{geometry}			
\usepackage[parfill]{parskip}				
\usepackage{graphicx}
\usepackage{amssymb}
\usepackage{epstopdf}
\title{A symmetrical theory of nonrelativistic quantum mechanics}
\author{Michael B. Heaney\\3182 Stelling Drive\\Palo Alto, CA 94303\\mheaney@alum.mit.edu}
\date{20 October 2013}				
\begin{document}
\maketitle
\begin{abstract}
This paper presents a new Symmetrical Theory (ST) of nonrelativistic quantum mechanics which postulates: quantum mechanics is a theory about complete experiments, not particles; a complete experiment is maximally described by a complex transition amplitude density; and this transition amplitude density never collapses. This new ST is compared to the Conventional Theory (CT) of nonrelativistic quantum mechanics for the analysis of a beam-splitter experiment. The ST makes several experimentally testable predictions that differ from the CT, which can be checked using existing technology. The ST also solves one part of the CT measurement problem, and resolves some of the paradoxes of the CT. This nonrelativistic ST is the low energy limit of a relativistic ST presented in an earlier paper \cite{Heaney1}.
\end{abstract}
\section{Introduction}
At the fifth Solvay Congress in 1927, Einstein presented a thought experiment to illustrate what he saw as a flaw in the Conventional Theory (CT) of quantum mechanics \cite{Valentini1}. Figure 1 shows a variation of his thought experiment. A single particle is released from a source $S$, travels to a beam-splitter $B$, then is captured by either detector $D1$ or detector $D2$. The particle is known to be localized at the source immediately before release, as shown in Fig. 2(a). The CT says that, after release, the particle's probability density evolves continuously and deterministically, splitting into two parts at the beam-splitter, as shown in Figs. 2(b) and 2(c). The CT says these two parts travel to detectors $D1$ and $D2$, as shown in Fig. 2(d). Upon measurements at the detectors, the CT says the probability density shown in Fig. 2(d) undergoes an instantaneous, indeterministic, and time-asymmetric collapse into either the probability density shown in Fig. 2(e) or the probability density shown in Fig. 2(f). This coordinated wavefunction collapse happens even if the two detectors are separated by a space-like interval at the time of collapse, suggesting that something is travelling faster than the speed of light. The instantaneous nature of this collapse also defines a privileged reference frame, which contradicts the special theory of relativity. Einstein believed this behavior was unphysical, which meant the CT was an incomplete theory. The still unresolved question of how (or if) collapse occurs is one part of the measurement problem of the CT \cite{WheelerZurek1}. Einstein suggested that some new mechanism, not described by the CT, was needed to replace wavefunction collapse. This paper presents a new Symmetrical Theory (ST) of nonrelativistic quantum mechanics that has such a mechanism, solving this part of the measurement problem. This ST also makes several experimentally testable predictions that differ from the CT, and resolves some of the paradoxes of the CT. 
\begin{figure}[htbp]
\begin{center}
\includegraphics[width=6.5in]{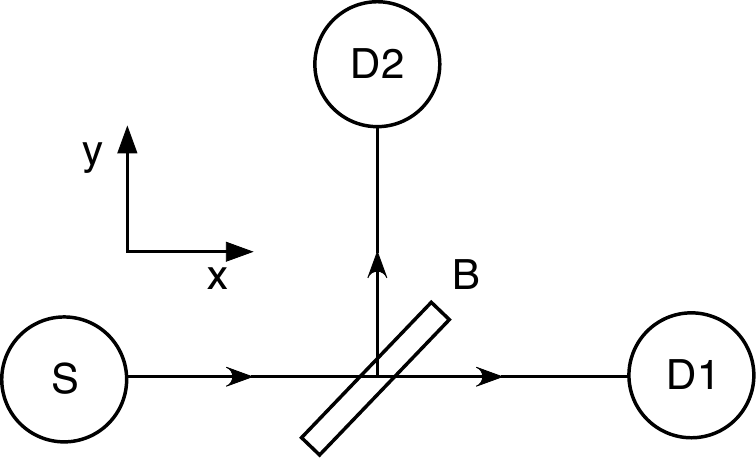}
\caption{The beam-splitter experiment. The source $S$ emits a single particle, which travels to the beam-splitter $B$. The particle is later found either in detector $D1$ or in detector $D2$. For an ensemble of identical initial conditions and a balanced beam-splitter, the probability of being found in each detector is 50\%.}
\label{default2}
\end{center}
\end{figure}
\section{Relation to Prior Work}
In 1955-56 Watanabe developed a time-symmetric formulation of the Conventional Theory (CT) of quantum mechanics by introducing a double inferential state-vector formalism (DIVF) \cite{Watanabe1,Watanabe2}. In the DIVF an isolated, individual particle is described by two state vectors: a predictive state vector $\vert\psi_p(t)\rangle$ that evolves forwards in time from the initial conditions according to the retarded Schr\"odinger equation, and a retrodictive state vector $\langle\psi_r(t)\vert$ that evolves backwards in time from the final conditions according to the advanced Schr\"odinger equation. The retarded Schr\"odinger equation (RSE) is just the usual Schr\"odinger equation, while the advanced Schr\"odinger equation (ASE) is the complex conjugate of the RSE. The DIVF allows one the freedom to impose independent initial \textit{and} final conditions on the evolution of a quantum system. Some parts of Watanabe's formulation were rediscovered by Aharonov, Bergmann, and Lebowitz in 1964 \cite{ABL}, and later renamed the two-state vector formalism (TSVF) \cite{AV, APT}. Various other time-symmetric formulations of quantum mechanics were developed by Davidon \cite{Davidon1}, Costa de Beauregard \cite{Beauregard1}, Roberts \cite{Roberts1}, Cramer \cite{Cramer1}, Hokkyo \cite{Hokkyo1,Hokkyo2}, Sutherland \cite{Sutherland1}, Pegg and Barnett \cite{PeggBarnett1}, Wharton \cite{Wharton1,Wharton2}, Miller \cite{Miller1}, and Gammelmark et al. \cite{Gammelmark1}.

Watanabe's DIVF is completely equivalent to the CT: \textquotedblleft Although some new points of view and a new formalism are introduced, the content of this paper will remain perfectly faithful to the accepted premises of classical and quantum physics \cite{Watanabe1}."  The TSVF reincarnation of the DIVF is also completely equivalent to the CT: \textquotedblleft To be clear: The phenomena under discussion and the questions they raise are new, but they can be fully addressed in standard quantum mechanics. We have not modified quantum mechanics by an iota, nor have we tinkered with the notion of time \cite{APT}." 

In contrast, the new Symmetrical Theory (ST) proposed in this paper and an earlier paper \cite{Heaney1} is a fundamentally different theory than the CT and the time-symmetric formulations described above. As a first difference, an implicit postulate of the CT and these time-symmetric formulations is that quantum mechanics is a theory about an isolated, individual particle (or a group of such particles). In contrast, the ST explicitly postulates that quantum mechanics is \textit{not} a theory about an isolated, individual particle (nor a group of such particles). Instead, the ST postulates that quantum mechanics is a theory about a \textquotedblleft complete experiment," defined as an isolated, individual particle (or a group of such particles) that starts with maximally specified initial conditions, evolves in space-time, then ends with maximally specified final conditions. This idea has been suggested before, in the context of the CT \cite{Bohr1,Feynman1}, but not in the context of a time-symmetric theory. As a second difference, the TSVF postulates that a \textquotedblleft particle" is completely described by a \textquotedblleft two-state vector," written as $\langle\phi\vert\thickspace\vert\psi\rangle$. This \textquotedblleft two-state vector" is not mathematically defined. In contrast, the ST postulates that a \textquotedblleft complete experiment" involving a particle is completely described by a complex transition amplitude density $\phi^\ast\psi$. The ST defines this transition amplitude density as the algebraic product of the two wavefunctions, which is a dynamical function of position and time. As a third difference, the CT and TSVF postulate that wavefunctions collapse upon measurement \cite{ACE1}, while the ST postulates that wavefunctions never collapse. Other fundamental differences will be described below, in a comparison of the CT and ST analyses of a beam-splitter experiment. 
\section{The Conventional Theory}
To understand the origins of and differences between the Conventional Theory (CT) and the new Symmetrical Theory (ST), it is necessary to start with the relativistic versions of these two theories, then take the nonrelativistic limits. This section will do this for the CT, while a later section will do this for the ST. 

The CT implicitly postulates that relativistic quantum mechanics is a theory about isolated, individual physical systems, such as a single subatomic particle \cite{Wachter1,Greiner1,Sakurai1,Bjorken1}. The CT explicitly postulates that such a system is maximally described by a wavefunction $\Theta(\vec{r},t)$, together with maximally specified initial conditions. The CT also postulates that the wavefunction $\Theta(\vec{r},t)$ of a free spin-0 particle of mass $m$ either evolves continuously and deterministically according to the relativistic Klein-Gordon equation (KGE):

\begin{equation}
\left(\frac{1}{c^2}\frac{\partial^2}{\partial t^2}-\nabla^2+\frac{m^2c^2}{\hbar^2}\right)\Theta=0,
\label{ }
\end{equation}

or undergoes an instantaneous, indeterministic, and time-asymmetric collapse into a different wavefunction $\Xi(\vec{r},t)$, upon measurement at time $t_f$:

\begin{equation}
\Theta(\vec{r},t_f)\longrightarrow\Xi(\vec{r},t_f).
\label{ }
\end{equation}

After this collapse, the CT assumes the new wavefunction $\Xi(\vec{r},t)$ evolves continuously and deterministically according to the KGE. 

The KGE has both positive energy $E_+$ and negative energy $E_-$ solutions, with energies:

\begin{equation}
E_\pm=\pm\sqrt{p^2c^2+m^2c^4}.
\label{ }
\end{equation}

The CT assumes the general solution $\Theta(\vec{r},t)$ of the KGE is a sum of a positive energy, retarded wavefunction $\Psi(\vec{r},t)$ and a negative energy, advanced wavefunction $\Phi^\ast(\vec{r},t)$:

\begin{equation}
\Theta = \Psi + \Phi^\ast.
\label{ }
\end{equation}

When taking the nonrelativistic limit of the KGE, the CT interprets $\Psi(\vec{r},t)$ as the nonrelativistic wavefunction of the particle. The CT either ignores $\Phi^\ast(\vec{r},t)$ or assumes it can be reinterpreted as the positive energy, retarded wavefunction of the associated antiparticle. In the nonrelativistic limit, the positive energy $E_+$ can be approximated by: 

\begin{equation}
E_+\approx E'+mc^2,
\label{ }
\end{equation}

where $0<E'\ll mc^2$. 
The CT assumes that, in the nonrelativistic limit, the positive energy solution $\Psi(\vec r,t)$ of the KGE can be approximated by:

\begin{equation}
\Psi(\vec r,t)\approx\psi(\vec r,t)\exp\left[\frac{-imc^2t}{\hbar}\right],
\label{ }
\end{equation}

where the time dependence of $\psi(\vec r,t)$ is:

\begin{equation}
\psi(\vec r,t)\propto\exp\left[\frac{-iE't}{\hbar}\right].
\label{ }
\end{equation}

Inserting Eq.(6) into Eq.(1), dropping a term proportional to $(E'/mc^2)^2$, and suppressing the rapidly oscillating term $\exp[-imc^2t/\hbar]$ gives the retarded Schr\"odinger equation (RSE) for a free particle:

\begin{equation}
i\hbar\frac{\partial\psi}{\partial t}=-\frac{\hbar^2}{2m}\nabla^2\psi.
\label{ }
\end{equation}

The CT implicitly postulates that nonrelativistic quantum mechanics is a theory about isolated, individual physical systems, such as a single subatomic particle \cite{Griffiths1,CTDL1,Shankar1,Dirac1}. The CT explicitly postulates that such a system is maximally described by the wavefunction $\psi(\vec{r},t)$, together with maximally specified initial conditions. The CT also explicitly postulates that the wavefunction $\psi(\vec{r},t)$ either evolves continuously and deterministically according to the RSE, or undergoes an instantaneous, indeterministic, and time-asymmetric collapse into a different wavefunction $\xi(\vec{r},t)$, upon measurement at time $t_f$:

\begin{equation}
\psi(\vec{r},t_f)\longrightarrow\xi(\vec{r},t_f).
\label{ }
\end{equation}

If we multiply Eq.(8) on the left by $\psi^\ast(\vec{r},t)$, multiply the complex conjugate of Eq.(8) on the left by $\psi(\vec{r},t)$, then take the difference of the two resulting equations, we get the CT local conservation law:

\begin{equation}
\frac{\partial\rho}{\partial t}+\nabla\cdot\vec{j}=0,
\label{ }
\end{equation}

where

\begin{equation}
\rho\equiv\psi^\ast\psi,
\label{ }
\end{equation}

and

\begin{equation}
\vec{j}\equiv\frac{\hbar}{2mi}\left(\psi^\ast\nabla\psi-\psi\nabla\psi^\ast\right).
\label{ }
\end{equation}

The CT interprets $\rho(\vec{r},t)$ as the probability density for finding the particle, and $\vec{j}(\vec{r},t)$ as the probability density current. These equations are unchanged if we substitute the complete wavefunction $\Psi(\vec r,t)$ for $\psi(\vec r,t)$.

If we assume $\psi(\vec{r},t)$ is normalized and goes to zero at $\vec{r}=\pm\infty$, and integrate the CT local conservation law over all space $V$, we get the CT global conservation law:

\begin{equation}
\iiint_{-\infty}^{+\infty}dV\rho(\vec{r},t)=1.
\label{ }
\end{equation}

This says the probability of finding the particle somewhere in space is always equal to one.
\begin{figure}[htbp]
\begin{center}
\includegraphics[width=6.5in]{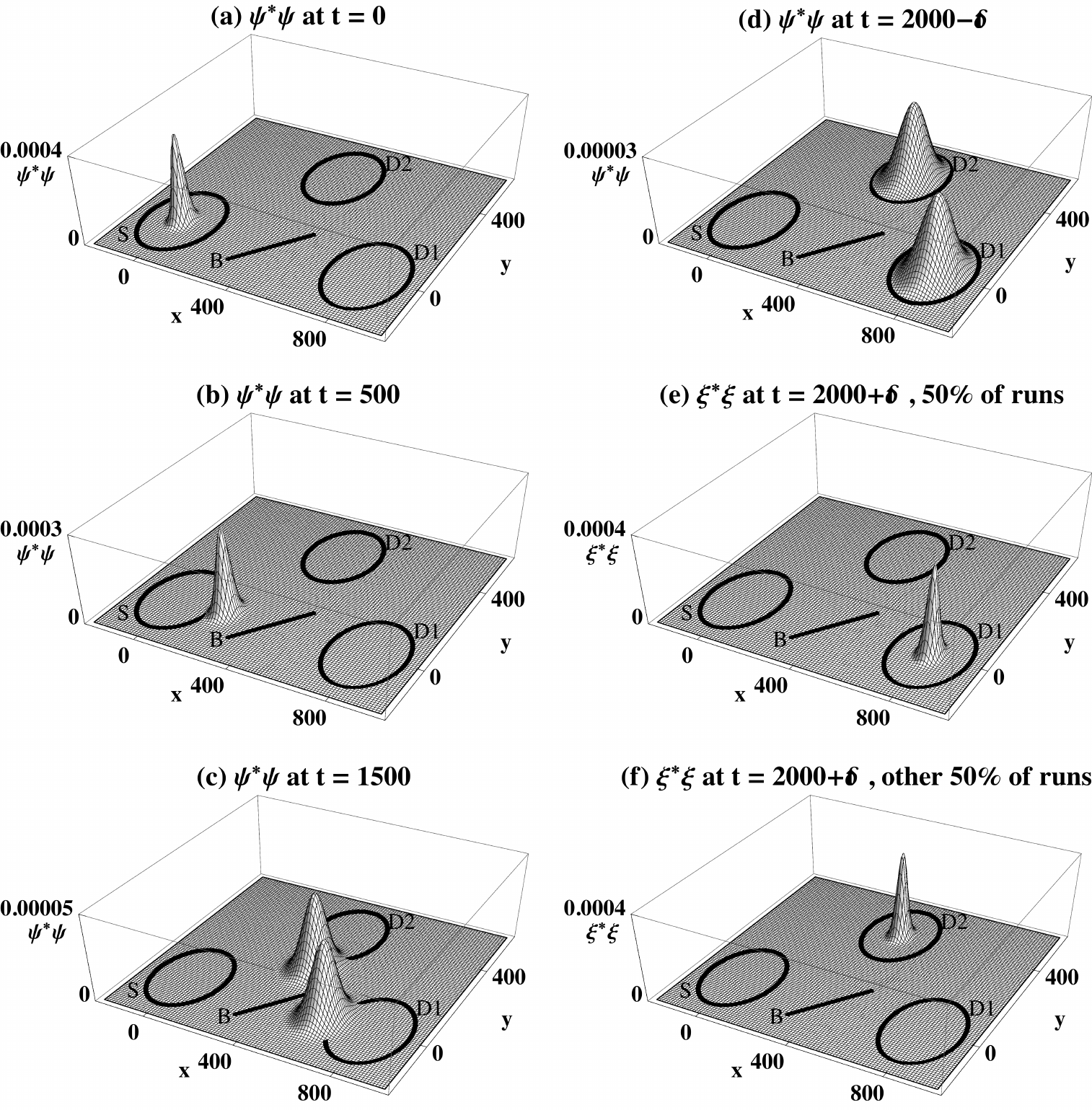}
\caption{The Conventional Theory (CT) analysis of the beam-splitter experiment. (a) The particle probability density $\psi^\ast\psi$ is initially localized at $S$. (b) It travels towards $B$. (c) It is split in two by $B$. The transmitted part travels towards $D1$, while the reflected part travels towards $D2$. (d) The two parts arrive in the detectors simultaneously, just before a measurement is made. (e) For 50\% of experiments, the probability density collapses into a localized and normalized probability density at $D1$ upon measurement. (f) For the other 50\% of experiments, the probability density collapses into a localized and normalized probability density at $D2$ upon measurement.}
\label{default2}
\end{center}
\end{figure}
\section{The Conventional Theory analysis of the beam-splitter experiment}
Now we will describe how the CT of nonrelativistic quantum mechanics analyzes the beam-splitter experiment shown in Fig. 1. The source $S$ is centered at $(x,y)=(0,0)$, the beam-splitter $B$ is at $(400,0)$, the detector $D1$ is at $(800,0)$, and the detector $D2$ is at $(400,400)$. The experiment is designed so that every particle that leaves the source $S$ will later be found at either detector $D1$ or detector $D2$, with equal probabilities. Let us use natural units and assume the following: a two-dimensional experiment; a particle mass $m\equiv1$; a positive energy, retarded wavefunction solution $\psi(\vec{r},t)$ to the RSE; and an initial normalized CT wavefunction $\psi(\vec{r},0)$ at the source that is a traveling gaussian with mean $(x_i,y_i)=(0,0)$, standard deviation $\sigma_i=20$, and momentum $k_x=0.4$. Figure 2(a) shows the particle's normalized CT probability density $\psi^\ast\psi$ at the initial time $t = 0$. It is localized at the source. Figure 2(b) shows the CT probability density at $t = 500$, halfway between the source and the beam-splitter. The probability density becomes less localized as time passes. Figure 2(c) shows the CT probability density at $t = 1500$, after interaction with the beam-splitter. One part of the wavefunction was transmitted through the beam-splitter and travels towards $D1$, while the other part of the wavefunction was reflected by the beam-splitter and travels towards $D2$. Figure 2(d) shows the CT probability density inside the detectors, immediately before detection at $t_f=2000$. Between $t = 0$ and $t = 2000 -\delta $, the CT probability density evolves continuously and deterministically, becoming progressively more delocalized as time passes. The volume under the curve $\psi^\ast\psi$ is conserved and equal to one. Figure 2(e) shows the CT probability density if the particle is detected at $D1$ at $t_f=2000$. The CT probability density has collapsed into a different probability density, localized at $D1$. This occurs for 50\% of experiments. Figure 2(f) shows the CT probability density if the particle is detected at $D2$ at $t_f=2000$. The CT probability density has collapsed into a different probability density, localized at $D2$. This occurs for the other 50\% of experiments. 

The CT measurement postulate says that, upon measurement at time $ t_f=2000$, the amplitude $A$ for a particle with wavefunction $\psi(\vec{r},t_f)$ to be measured as having the different wavefunction $\xi(\vec{r},t_f)$ is:

\begin{equation}
A\equiv \iiint_{-\infty}^{+\infty}dV\xi^\ast(\vec{r},t_f)\psi(\vec{r},t_f).
\label{ }
\end{equation}

The complex number $A$ is known as the CT transition amplitude. Let us assume the final CT wavefunction $\xi(\vec{r},t_f)$ is a traveling gaussian with the same properties as the initial CT wavefunction but localized at the center of $D1$, with the probability density shown in Fig. 2(e). Then numerical calculations predict $A=-0.19+0.40i$. The CT postulates that the probability $P$ for this transition is:

\begin{equation}
P\equiv A^\ast A.
\label{ }
\end{equation}

The positive real number $P$ is known as the CT transition probability. For the specific experiment shown in Fig. 2, numerical calculations predict $P=0.20$. (Note that this is not $0.50$ because we have specified a particular narrow final wavefunction inside the detectors.) If we instead assume the final CT wavefunction $\xi(\vec{r},t_f)$ is a traveling gaussian with the same properties as the initial CT wavefunction but localized at the center of $D2$ and having momentum $k_y=0.4$, with the probability density shown in Fig. 2(f), we get the same transition amplitude and transition probability. 

The CT collapse postulate says that, upon measurement at $t_f$, the wavefunction $\psi(\vec{r},t_f)$, and therefore the CT probability density of Fig. 2(d), undergoes an instantaneous, indeterministic, and time-asymmetric collapse into either the CT probability density of Fig. 2(e) or the CT probability density of Fig. 2(f). The CT interprets Eq.(14) as the wavefunction $\psi(\vec{r},t_f)$ collapsing into the different wavefunction $\xi(\vec{r},t_f)$ upon measurement at time $ t_f=2000$, with an amplitude equal to the overlap integral between the two wavefunctions at $ t_f$. The CT attributes the unpredictability of where the wavefunction collapses to fundamental quantum randomness. 
\section{The Symmetrical Theory}
To understand the origins of and differences between the nonrelativistic Conventional Theory (CT) and the new  nonrelativistic Symmetrical Theory (ST), it is necessary to start with the relativistic versions of these two theories, then take the nonrelativistic limits. This section will do this for the ST. The relativistic version of the ST was presented in an earlier paper \cite{Heaney1}.

The relativistic ST explicitly postulates that relativistic quantum mechanics is a theory about a \textquotedblleft complete experiment," defined as an isolated, individual particle (or a group of such particles) that starts with maximally specified initial conditions, evolves in space-time, then ends with maximally specified final conditions \cite{Heaney1}. The relativistic ST postulates that a \textquotedblleft complete experiment" is maximally described by a complex transition amplitude density composed of two wavefunctions: a positive energy, retarded wavefunction $\Psi(\vec{r},t)$, which satisfies the KGE and only the maximally specified initial conditions; and a negative energy, advanced wavefunction $\Phi^\ast(\vec{r},t)$, which satisfies the KGE and only the maximally specified final conditions. The relativistic ST postulates that this complex transition amplitude density always evolves continuously and deterministically, and never collapses. The nonrelativistic ST takes the nonrelativistic limit of the positive energy KGE to get the same retarded Schr\"odinger equation (RSE) and retarded wavefunction $\Psi(\vec{r},t)$ as the CT, but with a different interpretation. The nonrelativistic ST takes the nonrelativistic limit of the negative energy KGE to get the advanced Schr\"odinger equation (ASE) and the advanced wavefunction $\Phi^\ast(\vec{r},t)$ as follows: In the nonrelativistic limit, the negative energy $E_-$ solution of Eq. (3) can be approximated by:

\begin{equation}
E_-\approx-E'-mc^2,
\label{ }
\end{equation}

where $0<E'\ll mc^2$.

Now assume the negative energy solution of the KGE can be approximated by: 

\begin{equation}
\Phi^\ast(\vec r,t)\approx\phi^\ast(\vec r,t)\exp\left[\frac{imc^2t}{\hbar}\right],
\label{ }
\end{equation}

where the time dependence of $\phi^\ast(\vec r,t)$ is:

\begin{equation}
\phi^\ast(\vec r,t)\propto\exp\left[\frac{iE't}{\hbar}\right].
\label{ }
\end{equation}

Inserting Eq.(17) into Eq.(1), dropping a term proportional to $(E'/mc^2)^2$, and suppressing the rapidly oscillating term $\exp[imc^2t/\hbar]$ gives the advanced Schr\"odinger equation (ASE) for a free particle:

\begin{equation}
-i\hbar\frac{\partial\phi^\ast}{\partial t}=-\frac{\hbar^2}{2m}\nabla^2\phi^\ast.
\label{ }
\end{equation}

The ST explicitly postulates that nonrelativistic quantum mechanics is a theory about \textquotedblleft complete experiments," defined as isolated, individual physical systems that start with maximally specified initial conditions, evolve in space-time, then end with maximally specified final conditions. The ST postulates that such a complete experiment is maximally described by the algebraic product of two wavefunctions: a positive energy, retarded wavefunction $\psi(\vec{r},t)$, which satisfies the RSE and only the maximally specified initial conditions; and a negative energy, advanced wavefunction $\phi^\ast(\vec{r},t)$, which satisfies the ASE and only the maximally specified final conditions. The ST postulates that these two wavefunctions, and their algebraic product, always evolve continuously and deterministically, and wavefunction collapse never occurs. 

If we multiply Eq.(8) on the left by $\phi^\ast(\vec{r},t)$, multiply Eq.(19) on the left by $\psi(\vec{r},t)$, then take the difference of the two resulting equations, we get the ST local conservation law:

\begin{equation}
\frac{\partial\rho_s}{\partial t}+\nabla\cdot\vec{j_s}=0,
\label{ }
\end{equation}

where

\begin{equation}
\rho_s\equiv\phi^\ast\psi,
\label{ }
\end{equation}

and

\begin{equation}
\vec{j_s}\equiv\frac{\hbar}{2mi}\left(\phi^\ast\nabla\psi-\psi\nabla\phi^\ast\right).
\label{ }
\end{equation}

Note that $\rho_s(\vec{r},t)$ and $\vec{j_s}(\vec{r},t)$ are generally complex functions, and cannot be interpreted as probabilities. Instead, we will interpret $\rho_s(\vec{r},t)$ as the ST amplitude density for a complete experiment, and $\vec{j_s}(\vec{r},t)$ as the ST amplitude density current for a complete experiment. These equations are unchanged if we substitute the complete wavefunctions $\Psi(\vec r,t)$ for $\psi(\vec r,t)$ and $\Phi^\ast(\vec r,t)$ for $\phi^\ast(\vec r,t)$. 
 
If we assume $\psi(\vec{r},t)$ and $\phi^\ast(\vec{r},t)$ are each normalized and go to zero at $\vec{r}=\pm\infty$, and integrate the ST local conservation law over all space $V$, we get the ST global conservation law:

\begin{equation}
A_s\equiv   \iiint_{-\infty}^{+\infty}dV\phi^\ast(\vec r,t)\psi(\vec{r},t)   \equiv\iiint_{-\infty}^{+\infty}dV\rho_s(\vec{r},t),
\label{ }
\end{equation}

where the ST amplitude $A_s$ is generally a complex constant. The ST interprets $A_s$ as the amplitude for a complete experiment, and $P_s\equiv A_s^\ast A_s$ as the probability for a complete experiment. Equation (23) says the volume under the curve of the real part of $\rho_s(\vec{r},t)$ is conserved, and the volume under the curve of the imaginary part of $\rho_s(\vec{r},t)$ is also conserved. Then the probability for a complete experiment $P_s$ is always real and conserved, but generally not equal to one.

Note that:

\begin{equation}
P_s \equiv \iiint_{-\infty}^{+\infty}dV\rho^\ast_s(\vec{r},t) \iiint_{-\infty}^{+\infty}dV\rho_s(\vec{r},t) \neq \iiint_{-\infty}^{+\infty}dV\rho^\ast_s(\vec{r},t)\rho_s(\vec{r},t),
\label{ }
\end{equation}

which implies the ST does not predict the probability density for a particle to be found at any time between the initial conditions and the final conditions of a complete experiment. This is why $\rho^\ast_s\rho_s$ is not plotted in any of the figures. This is consistent with the ST postulate that quantum mechanics is a theory which describes complete experiments, not individual particles. 

In contrast to the CT, the ST postulates that the algebraic product $\phi^\ast(\vec r,t)\psi(\vec{r},t)$ varies deterministically and continuously for all times, and never undergoes collapse.
This is reasonable because the ST amplitude density automatically and smoothly reconverges to a distribution localized in the detector. This also means the integral of Eq.(23) can be evaluated at any time $t$.
\section{The Symmetrical Theory analysis of the beam-splitter experiment}
Now we will describe how the nonrelativistic ST analyzes the beam-splitter experiment shown in Fig.(1). Let us use natural units and assume the following: a two-dimensional experiment; a particle mass $m\equiv1$; a positive energy, retarded wavefunction solution $\psi(\vec{r},t)$ to the RSE; an initial normalized wavefunction $\psi(\vec{r},0)$ at the source that is a traveling gaussian with mean $(x_i,y_i)=(0,0)$, standard deviation $\sigma_i=20$, and momentum $k_x=0.4$; a negative energy, advanced wavefunction solution $\phi^\ast(\vec{r},t)$ to the ASE; and a final normalized gaussian wavefunction $\phi(\vec{r},2000)$ that is either centered at detector $D1$ with standard deviation $\sigma_i=20$ and momentum $k_x=0.4$, or centered at detector $D2$ with standard deviation $\sigma_i=20$ and momentum $k_y=0.4$. 

The ST transition amplitude density $\phi^\ast\psi$ is complex, making it difficult to plot on one graph and to compare with the real CT probability density $\psi^\ast\psi$. For these reasons, we will instead plot the absolute value of the ST transition amplitude density $|\phi^\ast\psi|$. Note that $|\phi^\ast\psi|$ does not obey the local and global conservation laws given by Eqs.(20) and (23), while $\phi^\ast\psi$ always obeys these conservation laws.

Figure 3 shows the general results for half of the complete experiments. The factor of $2^{-1/2}$ accounts for the fact that the transition probability is split evenly between arrival in $D1$ and arrival in $D2$. Figure 3(a) shows the absolute value of the ST amplitude density $2^{-1/2}|\phi^\ast\psi|$ at the initial time $t = 0$. It is localized at the source. Figure 3(b) shows $2^{-1/2}|\phi^\ast\psi|$ at $t = 500$, halfway between the source and the beam-splitter. Figure 3(c) shows $2^{-1/2}|\phi^\ast\psi|$ after interaction with the beam-splitter. All of the ST amplitude density was transmitted through the beam-splitter and travels towards $D1$. Figure 3(d) shows $2^{-1/2}|\phi^\ast\psi|$ immediately before detection at $t=2000$. Figure 3(e) shows $2^{-1/2}|\phi^\ast\psi|$ immediately after detection. It does not change after a measurement has been made: collapse never occurs. The predicted ST transition amplitude for the complete experiment shown in Fig. 3 is $A_s=-0.19+0.40i$. The ST says the probability $P$ for this transition is:

\begin{equation}
P_s\equiv A_s^\ast A_s.
\label{ }
\end{equation}

The positive real number $P_s$ is the ST complete experiment probability. For the specific experiment shown in Fig. 3, numerical calculations predict $P_s=0.20$. This reproduces one part of the CT prediction. (Note that this is not $0.50$ because we have specified a particular narrow final wavefunction inside the detectors.) 
\begin{figure}[htbp]
\begin{center}
\includegraphics[width=6.5in]{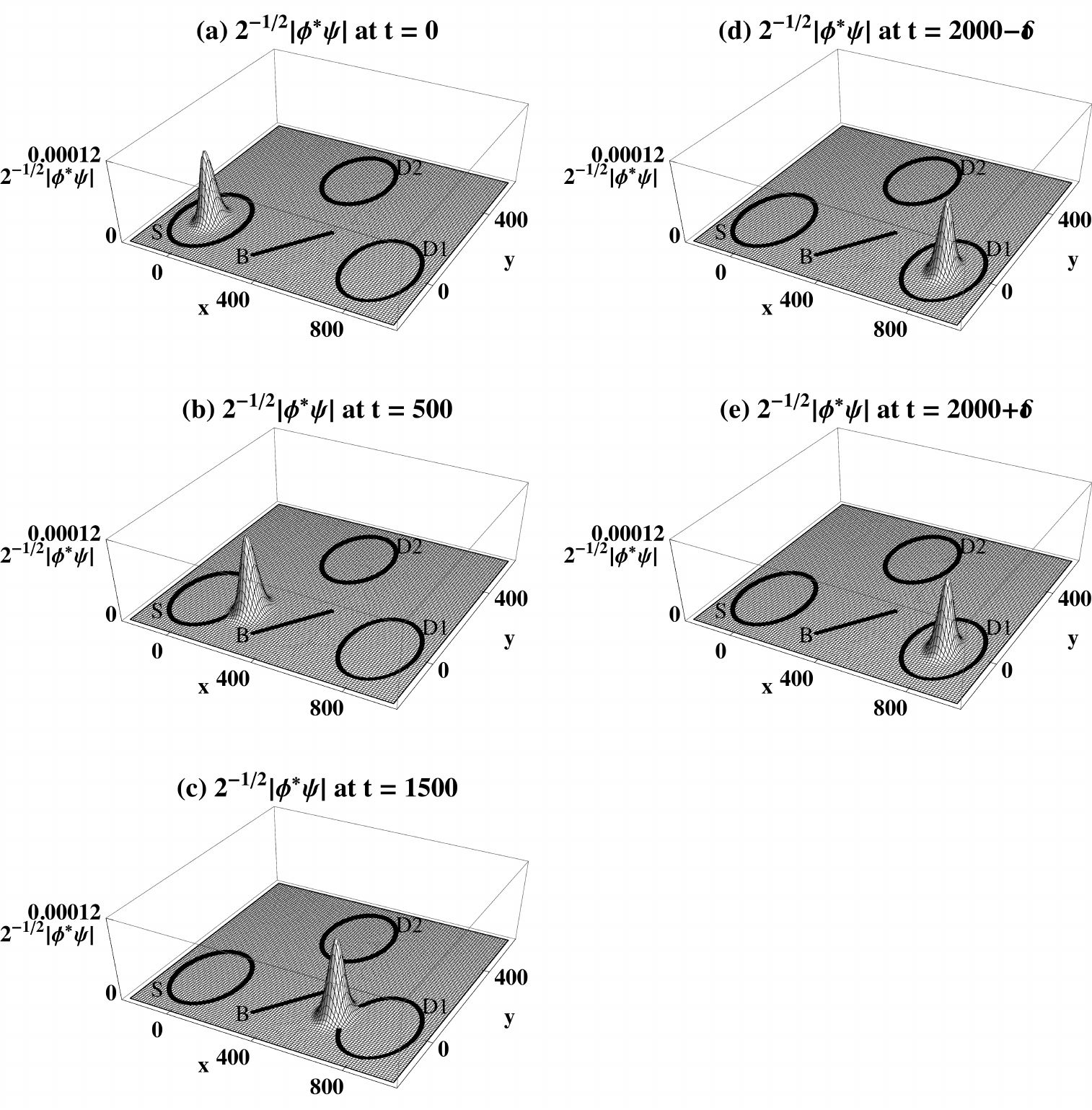}
\caption{The Symmetrical Theory (ST) analysis of the beam-splitter experiment, for 50\% of complete experiments: (a) The absolute value of the ST amplitude density $2^{-1/2}|\phi^\ast\psi|$ is initially localized at $S$. (b) It travels towards $B$. (c) All of it is transmitted through $B$ and travels towards $D1$. (d) It arrives at detector $D1$, just before a measurement is made at $t=2000$. (e) It does not change after a measurement has been made: collapse never occurs. (Note that $|\phi^\ast\psi|$ is shown for convenience of plotting and ease of comparison to $\psi^\ast\psi$, and only the complex transition amplitude $\phi^\ast\psi$ obeys local and global conservation laws.)}
\label{default2}
\end{center}
\end{figure}
\begin{figure}[htbp]
\begin{center}
\includegraphics[width=6.5in]{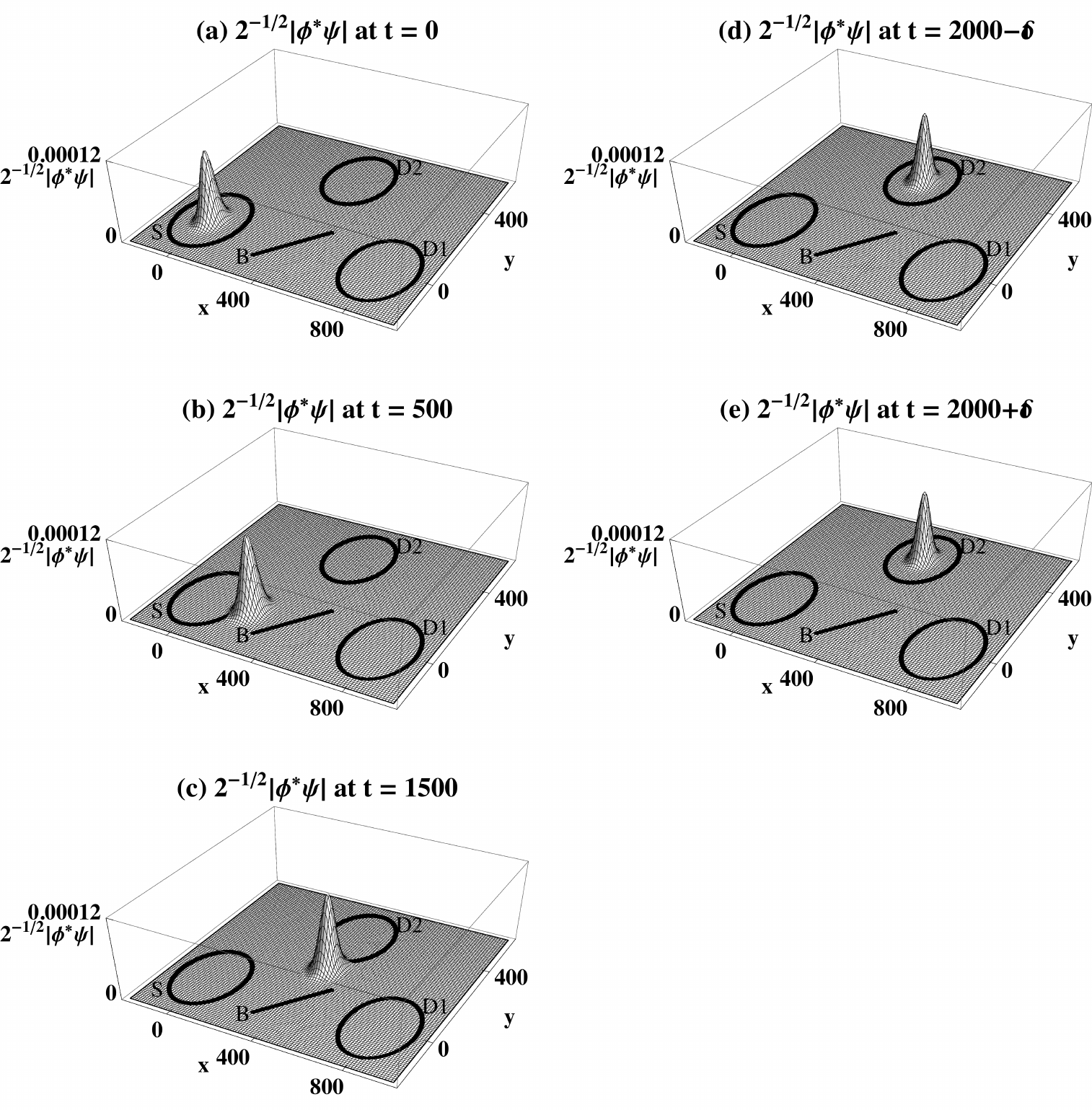}
\caption{The Symmetrical Theory (ST) analysis of the beam-splitter experiment, for the other 50\% of complete experiments: (a) The absolute value of the ST amplitude density $2^{-1/2}|\phi^\ast\psi|$ is initially localized at $S$. (b) It travels towards $B$. (c) All of it is reflected by $B$ and travels towards $D2$. (d) It arrives at detector $D2$, just before a measurement is made at $t=2000$. (e) It does not change after a measurement has been made: collapse never occurs. (Note that $|\phi^\ast\psi|$ is shown for convenience of plotting and ease of comparison to $\psi^\ast\psi$, and only the complex transition amplitude $\phi^\ast\psi$ obeys local and global conservation laws.)}
\label{default2}
\end{center}
\end{figure}

Figure 4 shows the general results for the other half of the complete experiments. The factor of $2^{-1/2}$ accounts for the fact that the transition probability is split evenly between arrival in $D1$ and arrival in $D2$. Figure 4(a) shows the absolute value of the ST amplitude density $2^{-1/2}|\phi^\ast\psi|$ at the initial time $t = 0$. It is localized at the source. Figure 4(b) shows $2^{-1/2}|\phi^\ast\psi|$ at $t = 500$, halfway between the source and the beam-splitter. Figure 4(c) shows $2^{-1/2}|\phi^\ast\psi|$ after interaction with the beam-splitter. All of the ST amplitude density was reflected by the beam-splitter and travels towards $D2$. Figure 4(d) shows $2^{-1/2}|\phi^\ast\psi|$ immediately before detection at $t=2000$. Figure 4(e) shows $2^{-1/2}|\phi^\ast\psi|$ immediately after detection. It does not change after a measurement has been made: collapse never occurs. The predicted ST transition amplitude and transition probability for the complete experiment shown in Fig. 4 are the same as for Fig. 3. This reproduces the other part of the CT prediction.

For this specific beam-splitter experiment, the CT and the ST give the same experimentally verifiable predictions. However, this does not imply that all of the experimentally verifiable predictions of the CT and the ST are the same, as will be discussed next.
\section{Experimental Tests to Distinguish the Two Theories} 
The CT predicts that the probability density $\psi^\ast\psi$ of a single particle splits into two parts at the beam-splitter, with one part travelling from $B$ to $D1$, while simultaneously the other part travels from $B$ to $D2$, as shown in Fig. 2(c). The CT also predicts that when these two parts arrive in detectors $D1$ and $D2$, they will randomly collapse into either a probability of one at $D1$ and a probability of zero at $D2$, as shown in Fig. 2(e), or a probability of zero at $D1$ and a probability of one at $D2$, as shown in Fig. 2(f). In contrast, the ST predicts the amplitude density $\phi^\ast\psi$ does not split into two parts at the beam-splitter, or collapse at the detectors: it either travels from $B$ to $D1$, as shown in Fig. 3(c), or it travels from $B$ to $D2$, as shown in Fig. 4(c). These conflicting predictions can be tested by \textquotedblleft weak" measurements of the presence of the densities on paths $B-D1$ and $B-D2$ for an ensemble of experiments \cite{Vaidman1}. If the \textquotedblleft weak" measurements show no correlation between the path taken by the  densities and the detector that measures the particle, the CT is correct. If the \textquotedblleft weak" measurements show a correlation between the path taken by the densities and the detector that measures the particle, the ST is correct. This experiment has not yet been done, but is feasible with existing technology.

The CT predicts that the width of the probability density $\psi^\ast\psi$ of a single particle increases monotonically between $S$ and the detectors. In contrast, the ST predicts that the width of the amplitude density $\phi^\ast\psi$ increases monotonically between $S$ and $B$, then decreases monotonically between $B$ and the detectors. These conflicting predictions can be tested by \textquotedblleft weak" measurements of the width of the distributions made halfway between $S$ and $B$ and halfway between $B$ and $D1$ for an ensemble of experiments. If the width of the distributions between $B$ and $D1$ is larger than the width of the distributions between $S$ and $B$, then the CT is correct. If the width of the distributions between $B$ and $D1$ is equal to the width of the distributions between $S$ and $B$, then the ST is correct. This experiment has not yet been done, but is feasible with existing technology.

The CT predicts that the shape of the probability density $\psi^\ast\psi$ of a single particle is always asymmetrical between $S$ and the detectors. In contrast, the ST predicts that the shape of the amplitude density $\phi^\ast\psi$ at $B$ is symmetrical. These conflicting predictions can be directly tested by \textquotedblleft weak" measurements of the shape of the distributions made at and near $B$ for an ensemble of experiments. If the shape of the distributions at $B$ is asymmetrical, then the CT is correct. If the shape of the distributions at $B$ is symmetrical, then the ST is correct. This experiment has not yet been done, but is feasible with existing technology.

Note that these experimental tests assume the experimental conditions defined earlier, where the initial and final measured distributions are assumed to be the same. This requires preselection and postselection of such experiments from an ensemble. 
\section{General Arguments for the Symmetrical Theory} 
The CT probability density $\psi^\ast\psi$ undergoes collapse upon measurement, which implies that measurements and observers play a special role in quantum mechanics \cite{vonNeumann1}. The ST amplitude density $\phi^\ast\psi$ never undergoes collapse, which implies that measurements and observers play no special role in quantum mechanics. This means the ST does not have one part of the \textquotedblleft measurement problem" which has plagued the CT since it's beginning \cite{WheelerZurek1}.

The CT postulates that an individual particle is maximally described by a retarded wavefunction and maximally specified initial conditions: the retarded wavefunction starts at an initial time, evolves only forwards in time, and collapses irreversibly and instantaneously at some undetermined future time. This means the CT is a \textquotedblleft presentist" theory, where only the present moment is real: the past is no longer real, and the future is not yet real. A \textquotedblleft presentist" theory is equivalent to a three-dimensional world, which changes as time passes. The ST postulates that a complete experiment is maximally described by the SI amplitude density $\phi^\ast\psi$, which is composed of a retarded wavefunction and an advanced wavefunction, and incorporates maximally specified initial and final conditions. This means the ST is an \textquotedblleft eternalist" theory, where the past, present, and future are equally real. The \textquotedblleft eternalist" theory is equivalent to a four-dimensional world, where time is just another parameter, like position. It is an experimental fact, proven by many experiments confirming the special theory of relativity, that the world is four-dimensional, not three-dimensional \cite{Petkov1}. In addition, the CT requires instantaneous wavefunction collapse, which defines a privileged reference frame. The ST has no such collapse, and no privileged reference frame. These facts mean the CT contradicts the special theory of relativity, while the ST does not. This is also true for the CT and ST of relativistic quantum mechanics \cite{Heaney1}. The \textquotedblleft presentist" nature of the CT is also inconsistent with the \textquotedblleft eternalist" nature of the general theory of relativity. The \textquotedblleft eternalist" nature of the ST may enable the unification of quantum mechanics and gravity.

The \textquotedblleft presentist" nature of the CT is the source of many paradoxical aspects of the CT interpretation of quantum mechanics. The \textquotedblleft eternalist" nature of the ST does not have such paradoxes. Consider Renninger negative-result experiments (also known as \textquotedblleft null" or \textquotedblleft interaction-free" measurements) \cite{IFM1,IFM2,IFM3}, and \textquotedblleft delayed-choice" experiments \cite{DC1, DC2,DC3}. In the simplest case of these experiments, a particle has two possible paths to travel from the source to the detector. The particle seems to know if one of those paths is (or will be) blocked or open, well before it has had time to reach the block. In the CT, this is unexplainable: how can the particle know what will happen in the future? In the ST, this is easily explained: a block in any space-time volume between the source and detector can prevent the advanced wavefunction from reaching the source at the time of emission, effectively letting the particle know that path is (or will be) blocked. 
 
\end{document}